\documentclass[aps,pra,showpacs,superscriptaddress,floatfix,nofootinbib]{revtex4}

\usepackage{amsmath} 
\usepackage{bbm}
\usepackage{epsfig}
\usepackage{times}
\makeatletter

\usepackage{psfrag,graphicx}
\usepackage{dcolumn}
\usepackage{amsmath,amssymb}
\usepackage{bm}
\usepackage[dvips]{color}
\usepackage[english]{babel}

\newcommand{\la}{\langle} \newcommand{\ra}{\rangle}

\newcommand{\be}{\begin{equation}}
\newcommand{\bea}{\begin{eqnarray}}
\newcommand{\eea}{\end{eqnarray}}
\newcommand{\ee}{\end{equation}}

\def\tr{\mathrm{Tr}}

\usepackage[T1]{fontenc}
\usepackage[latin9]{inputenc}
\usepackage{amsbsy}
\usepackage{graphicx}
\usepackage{amssymb}
\usepackage{esint}

\makeatletter
\numberwithin{equation}{section}
\numberwithin{figure}{section}

\usepackage{bbm}

\makeatother

\usepackage{babel}

\begin{document}

\title{Valence Bond States : Link models}

\author{E. Rico} 
\affiliation{Fakultat f\"ur Physik, Universit\"at Wien, Boltzmanngasse 5, A-1090 Vienna, Austria.}
\author{R. H\"ubener} 
\affiliation{Institut f\"ur Theoretische Physik, Universit\"at Innsbruck, Technikerstrasse 25, A-6020 Innsbruck, Austria.}  
\affiliation{Institut f\"ur Quantenoptik und Quanteninformation der \"Osterreichischen Akademie der Wissenschaften, Innsbruck, Austria}
\author{S. Montangero}
\affiliation{Institut f\"ur Quanteninformationsverarbeitung, Universit\"at Ulm, D-89069 Ulm, Germany}
\affiliation{NEST-CNR-INFM and Scuola Normale Superiore, Piazza dei Cavalieri 7, I-56126 Pisa, Italy}
\author{N. Moran}
\affiliation{Department of Mathematical Physics, National University of Ireland, Maynooth, Ireland}
\author{ B. Pirvu}
\affiliation{Fakultat f\"ur Physik, Universit\"at Wien, Boltzmanngasse 5, A-1090 Vienna, Austria.}
\author{J. Vala}
\affiliation{Department of Mathematical Physics, National University of Ireland, Maynooth, Ireland}
\affiliation{Dublin Institute for Advanced Studies, School of Theoretical Physics, 10 Burlington Rd, Dublin, Ireland}
\author{H.J. Briegel} 
\affiliation{Institut f\"ur Theoretische Physik, Universit\"at Innsbruck, Technikerstrasse 25, A-6020 Innsbruck, Austria.}  
\affiliation{Institut f\"ur Quantenoptik und Quanteninformation der \"Osterreichischen Akademie der Wissenschaften, Innsbruck, Austria}

\date{\today}

\begin{abstract}

An isotropic anti-ferromagnetic quantum state on a square lattice is characterized by symmetry arguments only. By construction, this quantum state is the result of an underlying valence bond structure without breaking any symmetry in the lattice or spin spaces. A detailed analysis of the correlations of the quantum state is given (using a mapping to a 2D classical statistical model and methods in field theory like mapping to the non-linear sigma model or bosonization techniques) as well as the results of numerical treatments (regarding exact diagonalization and variational methods). Finally, the physical relevance of the model is motivated. A comparison of the model to known anti-ferromagnetic Mott-Hubbard insulators is given by means of the two-point equal-time correlation function obtained i) numerically from the suggested state and ii) experimentally from neutron scattering on cuprates in the anti-ferromagnetic insulator phase.

\end{abstract}

\maketitle 

\section{Introduction}

The description of strongly correlated systems is one of the fundamental topics in condensed matter physics. These strongly correlated systems are studied mainly using low dimensional instances, where quantum phenomena are relevant and strong and thus one cannot expect to get an accurate understanding using classical concepts. Quantum phenomena contribute substantially in the Mott-Hubbard models, where the focus is on questions concerning quantum critical phenomena, metal-insulator transitions and quantum anti-ferromagnetism -- with obvious connections to quantum technology. Very helpful for the description of any physical system is the exploitation of symmetries. Using the symmetries, simplified models (toy models) can be built to understand some of the essentials of a given system.

In quantum anti-ferromagnetism, the traditional approach is linear spin wave theory (LSWT). Here, the assumption is that the quantum system is well described by taking the classical ground state (the N\'eel state) and a subsequent quantization of the (linearized) deviations from that state. In the last decades, new models of anti-ferromagnets have been studied to describe particular features of Mott systems where strong interaction effects cause a strong deviation from the picture of the N\'eel state. Among these new models, there are the quantum spin liquids, which almost appear to be the opposite of the classical N\'eel order. These spin liquids have at least two main characteristics that define them: (i) a spin liquid is a quantum state without magnetic long range order, and N\'eel order implies, in fact, magnetic long range order; (ii) a spin liquid is a quantum state without any spontaneously broken symmetry, and the N\'eel state breaks translational invariance and $SU(2)$ spin invariance.

These models are not just an academic exercise, they try to describe states of physical systems that can appear in Nature, like for instance in layers of $CuO_2$. Some of the appealing properties of these systems come from the "normal" state of the cuprate which is known \cite{Leggett:2006kq, Anderson:2004kl} to be (i) a spin singlet i.e., it does not break $SU(2)$ invariance and it does not have magnetic long range order; (ii) rotationally invariant, with a $d_{x^2-y^2}$ orbital symmetry; (iii) a time invariant quantum state. 

In the next sections, we will analyze a class of quantum states which have a local tensor description that enforces these important physical symmetries, the so called multipartite valence bond states (\emph{MVBS}). One of the most frequently studied states with such a local tensor structure is the ground state of the Affleck-Kennedy-Lieb-Tasaki (\emph{AKLT}) model \cite{Affleck:1987cm,Affleck:1987cy}, a spin-1 chain with a Heisenberg-like Hamiltonian. The purpose of this work is to extend the AKLT construction to higher dimensions and study the physical properties that emerge. The extension of the AKLT model that will be studied has been described and motivated in detail in a prior publication, see Ref.~\cite{Rico:2008rm}.

The paper is organized as follows: in Sec.~2, we describe the tensor structure underlying the construction of the ground state of the multipartite valence bond model. In that section, we introduce the notation and give a summary of the ideas behind the construction. The following sections are then dedicated to the study of the physical properties of the theoretically motivated model. To do so, we employ several different approaches and work out in detail what we infer and what relevance the results add to the theoretical model for the description of real-world physics. Along these lines, in Sec.~3, we use a mapping to a two dimensional statistical model to characterize the properties of the state. This mapping will allow us to extract the properties of the two dimensional quantum states from a one dimensional quantum system. In Sec.~4, 5 and 6, we give a detailed study of the correlations in the system applying well-known analytic tools (mapping to the non-linear sigma model, bosonization techniques) and numerical methods (exact diagonalization and variational methods). The main aim of these two sections is to try to discern between an algebraic or exponential decay of the equal time correlations in the multipartite valence bond state.

Finally, in Sec.~7, we give a physical motivation of the real-world relevance of our theoretical model. We compare the two-point equal-time correlation function obtained from theoretical predictions of our anti-ferromagnetic model and linear spin wave theory with data from neutron scattering experiments of cuprates in the anti-ferromagnet insulator phase.

\section{two-dimensional anti-ferromagnetic system. The model}

In this section, we characterize a quantum state defined on a 2D square lattice with symmetry arguments only. The aim of this section is to give an understandable summary of steps for the definition of the multipartite valence bond state but we will not try to give all the detailed ideas that brought us to the construction of the state, for that purpose we refer to \cite{Rico:2008rm}.

The two dimensional multipartite valence bond model is defined with the following requirements:
\begin{itemize}

\item It is a real singlet of $SU(2)$, i.e., it does not break $SU(2)$ invariance, it does not have magnetic long range order and it is a time invariant quantum state.

\item It is a homogeneous, translationally and rotationally invariant state, i.e., it is invariant under any point group symmetry of the lattice.

\item Its local degrees of freedom are characterized by superpositions of singlet (spin-0) and triplet (spin-1) representations of $SU(2)$.

\item It is the ground state of a local Heisenberg-like Hamiltonian.

\end{itemize}

Many of the features and structures that we will find in this work can be seen as a generalization of the \emph{AKLT} model \cite{Affleck:1987cm,Affleck:1987cy}. The basic building block of this ground state is given by a structure known as the \emph{valence bond}. We will hence call the resulting state the multipartite valence bond state (MPVBS).

\begin{figure}[!ht]
\begin{center}
\resizebox{!}{2.0cm}{\includegraphics{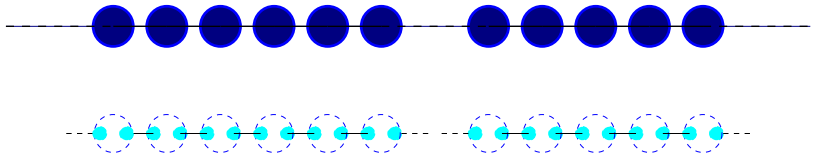}}
\caption[Valence bond ground state]{\label{aklt}(color online) Valence bond ground state. The first line with dark blue spots represents the physical state of the spin chain, while the second line represent the implementation of the state with two ancillae systems per site and a maximally entangled state between neighbor sites.}
\end{center}
\end{figure}

Following the original works \cite{Affleck:1987cm,Affleck:1987cy}, the \emph{AKLT} model describes a translationally invariant anti-ferromagnet spin-1 chain. Its construction is done in two steps:
\begin{enumerate}
\item In the ancilla picture, every other pair of contiguous spin-$\frac{1}{2}$ sites is projected into a singlet state, i.e., a maximally entangled state with zero total spin angular momentum (valence bond).
\item The remaining contiguous pairs are projected into the triplet subspace of two ancillary spin-$\frac{1}{2}$ subsystems, corresponding to a local spin-1 site of the "physical system".
\end{enumerate}
Denoting the state of a spin-$\frac{1}{2}$ ancilla subsystem by $|\alpha ) \in \mathbb{C}^2$, the first step in the construction of the \emph{VBS} is equivalent to fixing the state between neighboring ancilla spins to
\be
|0 ) =\sum_{\{\alpha,\beta\}=\{\uparrow,\downarrow\}} |\alpha) \epsilon_{\alpha \beta} |\beta ) = | \uparrow \downarrow ) - |\downarrow \uparrow ),
\ee
with $\epsilon_{\uparrow \downarrow}=-\epsilon_{\downarrow \uparrow}=1$ and $\epsilon_{\uparrow \uparrow}=-\epsilon_{\downarrow \downarrow}=0$. The projection of two spin-$\frac{1}{2}$ subsystems into the triplet subspace is imposed by the definition of the physical particle to be
\be
|\psi_{\alpha  \beta} \rangle = \frac{1}{\sqrt{2}} \left( |\alpha) |\beta) + |\beta) |\alpha) \right) = 
\begin{cases} 
\sqrt{2} |+1 \ra & \alpha = \beta = \uparrow \\
|0\ra & \alpha \neq \beta \\
\sqrt{2} |-1 \ra & \alpha = \beta = \downarrow.
\end{cases}
\ee

\begin{figure}[!ht]
\begin{center}
\resizebox{!}{3.0cm}{\includegraphics{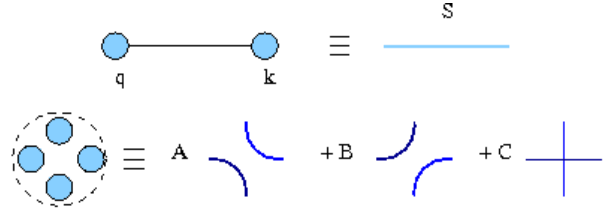}}
\caption{\label{ved} Representation of the physical Hilbert space, defined at every link by the projection of two spin-1/2 ancillae subsystems, and at every vertex by the projection of four spin-1/2 ancillae subsystems. In the analyzed example, a singlet state is chosen at the vertex (It turns out to be a special case of the classical six vertex model).}
\end{center}
\end{figure}

Following the valence bond construction of \emph{AKLT} \cite{Affleck:1987cm,Affleck:1987cy}, we will characterize a quantum system in a two-dimensional lattice \cite{Rico:2008rm}, but in a way that is  complementary to their extension of the model to two dimensions. We are looking for a uniform, translationally invariant singlet state in the square lattice. For this purpose, we restrict our attention to states of the following form (for graphical description see Fig. \ref{mvbs} and Fig. \ref{ved}):
\begin{figure}[!ht]
\begin{center}
\resizebox{!}{3.0cm}{\includegraphics{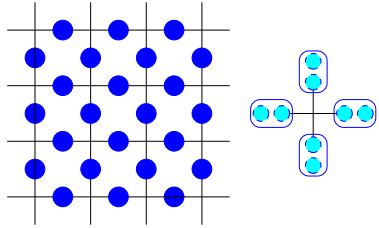}}
\caption[Multipartite valence bond state]{\label{mvbs} The multipartite valence bond ground state. On the left we find the lattice structure describing the model, where the local physical degrees of freedom are located at every link. The interactions take place around every vertex and involve the four nearest neighbor sites. On the right, the representation of the multipartite valence bond state in the ancillae picture.}
\end{center}
\end{figure}
\begin{enumerate}
\item On every link, the local Hilbert space is defined by a (weighted) projection of two ancillae spin-$1/2$ subsystems into the triplet and singlet representation of $SU(2)$.
\item On every vertex, four ancillae subsystems are projected into a singlet of $SU(2)$.
\end{enumerate}
Mathematically, both conditions are imposed using a set of appropriate tensors -- as in the case of the $AKLT$-model. First, at every link, the local quantum states are characterized by
\be
|\psi(\Lambda) \ra =a(0) \sigma^0 |0\ra + \sum_{s=\{x,y,z\}} a(s) \sigma^s ~  |s \ra =\frac{\sqrt{1+3\Lambda}}{2} \sigma^0 |0\ra +i \frac{\sqrt{1-\Lambda}}{2}  \sum_{s=\{x,y,z\}} \sigma^s ~  |s \ra,
\ee
where the state $|0\ra$ represents the singlet state and $\{|x\ra,|y\ra,|z\ra \}$ a basis of the triplet sector. The scalars $a(0)$, $\{a(s)\}$ give the amplitude of probability to find the state in the singlet and triplet sectors, respectively, such that: $a(x)=a(y)=a(z)$ are the probability amplitudes of the triplet sector; and $\sigma^s$ correspond to the usual Pauli matrices. 

Second, looking back at  the construction of \emph{AKLT}, they built a real singlet state out of real singlet valence bonds, mathematically, they linked every second ancillary spin-1/2 subsystem with a tensor $\Gamma_{a_1}^{b_2} = \epsilon_{a_1}^{b_2}$ that corresponds to the Levi-Civita tensor and is uniquely given by the symmetries of the problem. In our case, at the vertices, four spin-$1/2$ meet to form a singlet state of $SU(2)$. Decomposing the Hilbert space spanned by these spins into invariant subspaces of $SU(2)$, i.e., $\frac{1}{2} \otimes \frac{1}{2} \otimes \frac{1}{2} \otimes \frac{1}{2} = 0_+ \oplus 0_- \oplus 1 \oplus 2$, we find that the singlet sector is not unique but two-dimensional. If we rewrite the state at the vertex in a \emph{symmetric} gauge, we find that any real singlet state can be described by
\be
\Gamma_{a_1 c_3}^{b_2 d_4}[\phi]  =  \cos{\phi} \, \sigma^0_{a_1 d_4}  \sigma^0_{c_3 b_2} 
+  \frac{\sin{\phi}}{\sqrt{3}} \left( \sigma^x_{a_1 d_4}  \sigma^x_{c_3 b_2} + \sigma^y_{a_1 d_4}  \sigma^y_{c_3 b_2} + \sigma^z_{a_1 d_4}  \sigma^z_{c_3 b_2} \right)
\ee
where the angle $\phi$ determines a weight of the positive and the negative chiral singlet sectors \cite{Wen:1989ys}. There are two special points $\phi=\{ 0, \frac{\pi}{2} \} $ where this tensor describes a rotationally invariant real singlet.

It is straightforward to realize that the multipartite entangled state at every vertex corresponds to a special case of the classical six vertex model (see for instance \cite{Baxter:1989zr,Gomez:1996ly} and references therein), so that the state is decomposed in a classical structure ({\emph{scaffolding}}) defined at the vertex and the link where we place the singlet-triplet degrees of freedom. The six vertex model characterized by the tensor $\Gamma_{a_1 c_3}^{b_2 d_4}[\phi]$ lies on the critical line with anisotropic parameter $\Delta=1$. This critical line describes the phase transition between the disordered phase and the ferroelectric phase of the six vertex model.


\section{Ground state properties and correlations. Expectation values}

In this section, we characterize the quantum state we have just defined and obtain its correlations. Any expectation value can be calculated via a mapping of the 2D multipartite valence bond state to a 2D classical statistical model, and from there to a 1D quantum mechanical model. Hence, we obtain the properties of a two dimensional quantum model from the behavior of a one dimensional quantum system. In the following, we give the detailed steps for this mapping.

Given the 2D multipartite valence bond state, we map the (local) projectors to classical vertex matrices $R$. These matrices are hence derived directly from the tensors defining the projections at the links and at the vertices. E.g., in the calculation of the vacuum-to-vacuum expectation value, the projectors defining the valence bond model are mapped to two different classical structures:
\begin{itemize}
\item At every link, a transfer matrix is defined by $E=\sum_{s=\{0,x,y,z\}} |a(s)|^2 ~~ \left( \sigma^s \right)^* \otimes \sigma^s$
\item At every vertex, a vertex matrix is defined by $V=\Gamma^* \otimes \Gamma$
\end{itemize}
and following from these definitions, the resulting $R$-matrix is given by
\be
R^{i_2,i_4}_{i_1,i_3} = \sum_{\{j_1,j_2,j_3,j_4\}} \sqrt{E}_{i_1,j_1} \sqrt{E}_{i_3,j_3} V^{j_2,j_4}_{j_1,j_3} \sqrt{E}_{j_2,i_2} \sqrt{E}_{j_4,i_4}
\ee

The vacuum-to-vacuum expectation value in the valence bond state now corresponds to the calculation of the classical partition function of the derived classical vertex model:
\be
\la \psi | \psi \ra = \sum_{\text{all configurations}} \prod_{\text{lattice}} R^{ij}_{lk}   = \mathcal{Z}_{2D}
\ee
where we sum over all possible configurations compatible with the weights $R$.
\begin{figure}[!ht]
\begin{center}
\resizebox{!}{4.0cm}{\includegraphics{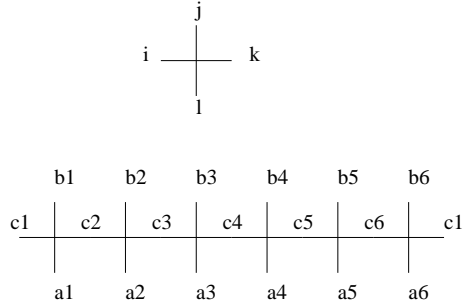}}
\caption{Vertex model in a classical context. The top part represents the vertex $R^{ij}_{lk} $, the Boltzmann weight of the vertex model. The bottom part represents the row-to-row transfer matrix for the vertex model.}
\end{center}
\end{figure}

The classical $R$-matrix or Boltzmann weight now depends on two parameters: $\phi$, the angle in the vertex state, and $\Lambda$, the difference in the amplitude of probability of being in the triplet and singlet sector at every link. Because we would like to interpret the $R$-matrix as the Boltzmann weight of a hermitian Hamiltonian, we have to look at the eigenvalues of this $R$-matrix and check for which values of the parameters they are positive definite. We restrict our attention, without loss of generality, to values of $\Lambda>0$. The eigenvalues of the $R$-matrix that can be negative are
\be
\begin{split}
r_1&=\frac{\Lambda}{3}\left(\sqrt{3} \sin{(2\phi)} - 3 \cos{(2 \phi)} \right)\\
r_2&=\frac{\Lambda^2}{3}\left(\sqrt{3} \sin{(2\phi)} - 3 \cos{(2 \phi)} \right).
\end{split}
\ee
In both cases, the $R$-matrix is positive definite if $\phi \in [\frac{\pi}{6}, \frac{2 \pi}{3}]$. Thus we restrict the study of the vertex model to the parameter regimen defined by $\Lambda \in [0,1]$ and $\phi \in [\frac{\pi}{6}, \frac{2 \pi}{3}]$.

From this point, we proceed with another well-known mapping, the equivalence between $D$-dimensional classical statistical models and $d=D-1$ dimensional quantum mechanical problems. In quantum mechanical problems, \emph{time} evolution is carried out by the row-to-row transfer matrix $T$, with matrix elements $T_{ab}$ given by
\be
T_{ab} = R^{c_1b_1}_{a_1c_2}  R^{c_2b_2}_{a_2c_3} \cdots R^{c_nb_n}_{a_nc_1}.
\ee
If we understand this time evolution operator as the exponential of a quantum Hamiltonian $\hat{H}_{1d}=- \log{T}$, then the classical partition function $\mathcal{Z}_{2D}$ is just given by the trace of the density matrix of a Gibbs ensemble $\mathcal{Z}_{2D}=\tr{\left( e^{-N \hat{H}_{1d}} \right)}$.

\begin{figure}[!ht]
\begin{center}
\resizebox{!}{3.0cm}{\includegraphics{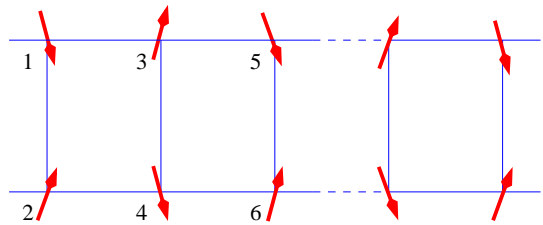}}
\caption{\label{ladder} Ladder model of spin-$1/2$ particles. This one dimensional quantum model describes the correlations of the two-dimensional multipartite quantum valence bond state. For more details, please, see the main text.}
\end{center}
\end{figure}

To write down the one dimensional Hamiltonian explicitly, we first realize that the indices of the $R$-matrix have rank four. The corresponding Hilbert space can be represented by two quantum spin-$1/2$ systems. The one dimensional quantum Hamiltonian can thus be described as a ladder of coupled spin-$1/2$ particles. To build the quantum ladder Hamiltonian we will use the spin operators
\be
S^x= \frac{1}{2} \begin{pmatrix} 0 & 1 \\ 1 & 0 \end{pmatrix}; ~~~~S^y= \frac{1}{2} \begin{pmatrix} 0 & -i \\ i & 0 \end{pmatrix}; ~~~~S^z= \frac{1}{2} \begin{pmatrix} 1 & 0 \\ 0 & -1 \end{pmatrix}; 
\ee
and we will use a notation such that the spins in the upper row are labelled with consecutive odd integers and the spins in the lower row are labelled with consecutive even integers. In summary, the ladder Hamiltonian can be written as
\be
\begin{split}
\label{hamlad}
H&=\beta_{1} \left( (\vec{S}_1\cdot \vec{S}_2) + (\vec{S}_3 \cdot \vec{S}_4) \right) +\beta_{2} \left( (\vec{S}_1 \cdot \vec{S}_3) + (\vec{S}_2 \cdot \vec{S}_4) \right) +\beta_{3} \left( (\vec{S}_1 \cdot \vec{S}_4) + (\vec{S}_2 \cdot \vec{S}_3) \right) +\\
&\beta_{4} \left( (\vec{S}_1\cdot  \vec{S}_2)  (\vec{S}_3 \cdot \vec{S}_4) \right) +\beta_{5} \left( (\vec{S}_1\cdot  \vec{S}_3)  (\vec{S}_2 \cdot \vec{S}_4) \right) + \beta_{6} \left( (\vec{S}_1 \cdot \vec{S}_4)  (\vec{S}_2 \cdot \vec{S}_3) \right), 
\end{split}
\ee
where the interaction involves the nearest four spin-$1/2$ in a plaquette of the spin ladder.

\begin{figure}[!ht]
\begin{center}
\resizebox{!}{6.0cm}{\includegraphics{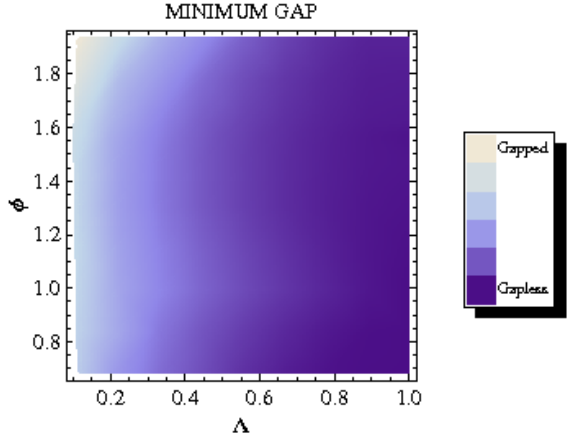}}
\caption{\label{dmrg} Minimum gap of the ladder Hamiltonian that encodes the behaviour of the correlation length of the 2D multipartite valence bond state in $\Lambda$ and $\phi$. This phase diagram is built with 100 sample points using a DMRG method. Dark color describes states with small energy gaps and light color bigger ones.}
\end{center}
\end{figure}

The values of the constants $\beta_i$ depend on two parameters $\Lambda$ and $\phi$. Initially, we will fix $\phi = \frac{\pi}{2}$, for two reasons: (i) for simplicity, in a preliminary study of the phase diagram over $\{ \Lambda \, , \,\phi \}$ and due to the fact that these values describe a rotationally invariant real singlet state. (ii) as we mentioned at the beginning of this paper, the different values of $\phi$ describe different instances of the six vertex model, nonetheless all of them correspond to the anisotropy parameter $\Delta=1$ and lie on the critical line of the six vertex model. (iii) Our numerical study of the phase diagram $\{ \Lambda$ and $\phi \}$ with the DMRG method \cite{White:1992fs,Schollwoeck:2004pz} shows that the parameter $\phi$ is irrelevant (see Fig.\ \ref{dmrg}).

Then, the constants $\beta_i(\Lambda, \frac{\pi}{2})$ are given by
\be
\begin{split}
\label{ladpar}
&\beta_1=\frac{1}{12 \sqrt{9+6\Lambda^2+49 \Lambda^4}} \\
&[ \left(  \sqrt{9+6\Lambda^2+49 \Lambda^4} \log{9} + 7 \Lambda^2 \log{36} - \log{46656} - 2 (3- 7 \Lambda^2 + 4\sqrt{9+6\Lambda^2+49 \Lambda^4} ) \log{\Lambda} \right) + \\
&+\left((-3 + 7 \Lambda^2) \log{(3 + 7 \Lambda^2 - \sqrt{9 + 6 \Lambda^2 + 49 \Lambda^4})} +  3 (3 - 7 \Lambda^2) \log{(3 + 7 \Lambda^2 + \sqrt{9 + 6 \Lambda^2 + 49 \Lambda^4})}
 \right) ]\\
&\beta_2= \frac{1}{12} \left( \log{(6561/\Lambda^4)} + \frac{2 ( \sqrt{9+6\Lambda^2+49 \Lambda^4} \log{\Lambda} + (-3 + \Lambda (12+ 7 \Lambda)) \log{(\frac{3 + 7 \Lambda^2 + \sqrt{9 + 6 \Lambda^2 + 49 \Lambda^4}}{6 \Lambda})}) }{ \sqrt{9+6\Lambda^2+49 \Lambda^4}} \right) \\
&\beta_3= \frac{1}{12} \left( \log{(9/\Lambda^4)} + \frac{2 ( \sqrt{9+6\Lambda^2+49 \Lambda^4} \log{\Lambda} + (-3 + \Lambda (-12+ 7 \Lambda)) \log{(\frac{3 + 7 \Lambda^2 + \sqrt{9 + 6 \Lambda^2 + 49 \Lambda^4}}{6 \Lambda})}) }{ \sqrt{9+6\Lambda^2+49 \Lambda^4}} \right)  \\
&\beta_4=\frac{1}{3 \sqrt{9+6\Lambda^2+49 \Lambda^4}} \\
&[ \left(-  \sqrt{9+6\Lambda^2+49 \Lambda^4} \log{9} - 7 \Lambda^2 \log{36} + \log{46656} - 2 (-3 + 7 \Lambda^2 + 2 \sqrt{9+6\Lambda^2+49 \Lambda^4} ) \log{\Lambda} \right) + \\
&+\left((3 - 7 \Lambda^2) \log{(3 + 7 \Lambda^2 - \sqrt{9 + 6 \Lambda^2 + 49 \Lambda^4})} +  3 (-3 + 7 \Lambda^2) \log{(3 + 7 \Lambda^2 + \sqrt{9 + 6 \Lambda^2 + 49 \Lambda^4})}
 \right) ]\\
&\beta_5=\frac{1}{3} \left( 2\log{(\Lambda)} + \frac{ ( \sqrt{9+6\Lambda^2+49 \Lambda^4} \log{81} - 2 (-3 + \Lambda (12+ 7 \Lambda)) \log{(\frac{3 + 7 \Lambda^2 + \sqrt{9 + 6 \Lambda^2 + 49 \Lambda^4}}{6 \Lambda})}) }{ \sqrt{9+6\Lambda^2+49 \Lambda^4}} \right)  \\
&\beta_6= \frac{1}{3} \left( \log{(\Lambda^4 / 9)} - \frac{2 ( \sqrt{9+6\Lambda^2+49 \Lambda^4} \log{\Lambda} + (-3 + \Lambda (-12+ 7 \Lambda)) \log{(\frac{3 + 7 \Lambda^2 + \sqrt{9 + 6 \Lambda^2 + 49 \Lambda^4}}{6 \Lambda})}) }{ \sqrt{9+6\Lambda^2+49 \Lambda^4}} \right) 
\end{split}
\ee

\begin{figure}[!ht]
\begin{center}
\resizebox{!}{4.0cm}{\includegraphics{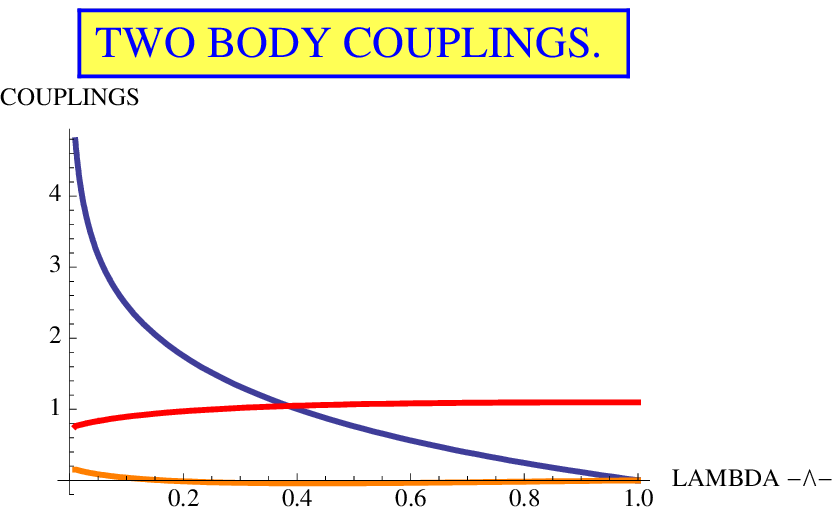}}
\caption{\label{onebody} Values of the coupling constant that involve two-body interactions. Blue corresponds to the coupling $\beta_1$ along the rung of the ladder; Red corresponds to the coupling $\beta_2$ parallel to the spin-$1/2$ chains; Orange corresponds to the coupling $\beta_3$ between diagonal spins in the plaquette.}
\end{center}
\end{figure}

\begin{figure}[!ht]
\begin{center}
\resizebox{!}{4.0cm}{\includegraphics{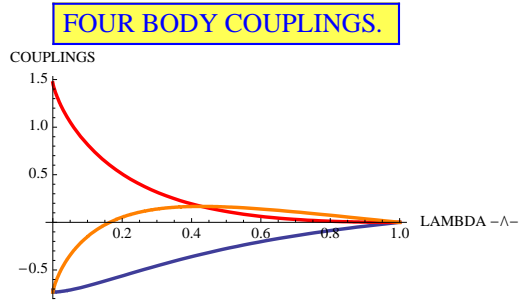}}
\caption{\label{twobody} Values of coupling constant that involve four body interactions. Blue corresponds to the coupling $\beta_4$ along the rung of the ladder; Red corresponds to the coupling $\beta_5$ parallel to the spin-$1/2$ chains; Orange corresponds to the coupling $\beta_6$ between diagonal spins in the plaquette.}
\end{center}
\end{figure}

It is illuminating to show a plot of the values of these coupling constants as a function of the parameter $\Lambda$ (see Fig. \ref{onebody} and \ref{twobody}).
From this first figure, Fig. \ref{onebody}, we can deduce that the coupling between diagonal spins in the plaquette is negligible; and when $\Lambda \to 1$ the ladder decouples into two independent Heisenberg spin-$1/2$ chains, since $\beta_2 \to 1$ and $\beta_1 \to 0$. This fact allows us to do perturbation theory around two weakly coupled Heisenberg chains \cite{Gogolin:1998qq}. 

From the second picture, Fig. \ref{twobody}, we can deduce that the values of the three coupling constants of the four spins in a plaquette are of the same order, i.e., none of them are negligible and there are regions where these couplings can be negative so that they cause ferromagnetic interaction.

\section{Semiclassical analysis: Mapping to the non-linear sigma model.}

We have seen that the expectation value of any operator in the 2D multipartite valence bond state can be mapped to a one dimensional quantum problem that is equivalent to a quantum ladder with two body and four body interactions. To get some insight into the behaviour of the system, we are going to use a semiclassical analysis using the Haldane mapping \cite{Haldane:1983tg,Affleck:1987uo,Brezin:1989qv} for the ladder problem \cite{Sierra:1996xe} (see also \cite{Martin-Delgado:1995pi}). This mapping is a commonly used tool in quasi-one dimensional quantum models which allows to get a first description of the excitations of these quantum models.

We should point out that although the picture that emerges with this semiclassical study follows some of the results that we will get with the numerical methods, the limits of validity of this approach should be analyzed in more detail. In fact, it is known that the presence of plaquette interactions in a ladder system gives a rich phase diagram with different critical points \cite{Shelton:1996la, Nersesyan:1997ek, Nersesyan:2003it, Mueller:2002zr, Lauchli:2003qq}. Also see \cite{Chakravarty:1988rr,Chakravarty:1989lq,Manousakis:1991dq} where this semiclassical approach is used to study two-dimensional anti-ferromagnets at low temperature.

Before entering in details, we fix the notation of the spin operator in the ladder $S^a_{\mu m}$, with $[S^a_{\mu m},S^b_{\nu n}] = i \delta_{\mu,\nu} \delta_{m,n} \sum_c \epsilon^{abc} S^c_{\mu m}$, in such a way that it carries three indices: $a \in \{x,y,z\}$ labels the direction of the spin operator; $\mu \in \{1,2\}$ labels the chain in the ladder and $m$ labels the position along the chain.

The semiclassical analysis starts with a mapping of the spin operator at every rung to a combination of a local staggered magnetization operator $\vec{\varphi}$ and a fluctuating one $\vec{l}$,
\be
\vec{S}_{\mu m} = \frac{1}{2} \vec{l}_m + (-1)^{m+\mu} s \vec{\varphi}_m; ~~~~~~ \vec{l}_m \cdot \vec{\varphi}_m=0
\ee
where $s$ is the value of the representation of the spin operator, i.e., $\left( \vec{S}_{\mu m} \right)^2 = s(s+1)$. With this definition and the commutation relation of the spin operators, it is straightforward to obtain
\be
\begin{split}
&[l^a_m,l^b_n] = i \delta_{n,m} \sum_c \epsilon^{abc} l^c_m;~~~~~~~~[l^a_m,\varphi^b_n] = i \delta_{n,m} \sum_c \epsilon^{abc} \varphi^c_m; \\
&[\varphi^a_m,\varphi^b_n] = i \frac{\delta_{n,m}}{s^2} \sum_c \epsilon^{abc} l^c_m \to 0, ~~ (s\gg 1); ~~ \left(\vec{ \varphi} _m \right)^2 = 1 + \frac{1}{s} - \frac{\left(\vec{l}_m\right)^2}{4 s^2} \to 1, ~~ (s\gg 1);
\end{split}
\ee
In the continuum limit, $\vec{\varphi}$ will become the field of the $O(3)$ non-linear $\sigma$-model and $\vec{l}$ will become the generator of rotations, i.e.\ $\vec{l} \simeq \vec{\varphi} \times \frac{d \vec{\varphi}}{d t}$

Hence, the set of operators that defined our Hamiltonian are mapped to
\be
\begin{split}
\left( \vec{S}_{1n} \cdot \vec{S}_{2n} \right) &= \frac{1}{2} \left( \vec{l}_n \right)^2 - s(s+1) \\
\left( \vec{S}_{1n} \cdot \vec{S}_{1,n+1} \right) &= \frac{1}{4} \vec{l}_n \cdot  \vec{l}_{n+1} + \frac{(-1)^n}{2} s \left( \vec{l}_n \cdot  \vec{\varphi}_{n+1}  - \vec{\varphi}_n \cdot  \vec{l}_{n+1} \right) - s^2  \vec{\varphi}_n \cdot  \vec{\varphi}_{n+1} \\
\left( \vec{S}_{2n} \cdot \vec{S}_{2,n+1} \right) &= \frac{1}{4} \vec{l}_n \cdot  \vec{l}_{n+1} - \frac{(-1)^n}{2} s \left( \vec{l}_n \cdot  \vec{\varphi}_{n+1}  - \vec{\varphi}_n \cdot  \vec{l}_{n+1} \right) - s^2  \vec{\varphi}_n \cdot  \vec{\varphi}_{n+1} \\
\left( \vec{S}_{1n} \cdot \vec{S}_{2,n+1} \right) &= \frac{1}{4} \vec{l}_n \cdot  \vec{l}_{n+1} - \frac{(-1)^n}{2} s \left( \vec{l}_n \cdot  \vec{\varphi}_{n+1}  + \vec{\varphi}_n \cdot  \vec{l}_{n+1} \right) + s^2  \vec{\varphi}_n \cdot  \vec{\varphi}_{n+1} \\
\left( \vec{S}_{2n} \cdot \vec{S}_{1,n+1} \right) &= \frac{1}{4} \vec{l}_n \cdot  \vec{l}_{n+1} + \frac{(-1)^n}{2} s \left( \vec{l}_n \cdot  \vec{\varphi}_{n+1}  + \vec{\varphi}_n \cdot  \vec{l}_{n+1} \right) + s^2  \vec{\varphi}_n \cdot  \vec{\varphi}_{n+1},
\end{split}
\ee

Assuming that  $\vec{\varphi}$ and $\vec{l}$ are slowly varying on the scale of the lattice, we calculate the Hamiltonian in a gradient expansion with a lattice spacing $\delta$. Keeping term up to $O\left( ( \frac{\partial \vec{\varphi}}{\partial x} )^2 \right)$ and $O(\vec{l}^2)$, only since, $\vec{l}$ effectively contain a time derivative and dropping constant terms, we obtain 
\be
\begin{split}
\left( \vec{S}_{1n} \cdot \vec{S}_{2n} \right) + \left( \vec{S}_{1,n+1} \cdot \vec{S}_{2,n+1} \right) &\to \left( \vec{l}(x) \right)^2  \\
\left( \vec{S}_{1n} \cdot \vec{S}_{1,n+1} \right) + \left( \vec{S}_{2n} \cdot \vec{S}_{2,n+1} \right) &\to \left( \vec{l}(x) \right)^2 + \frac{s^2 \delta^2}{2} \left( \frac{\partial \vec{\varphi}}{\partial x} \right)^2 \\
\left( \vec{S}_{1n} \cdot \vec{S}_{2,n+1} \right) + \left( \vec{S}_{2n} \cdot \vec{S}_{1,n+1} \right) &\to  - \frac{s^2 \delta^2}{2} \left( \frac{\partial \vec{\varphi}}{\partial x} \right)^2 \\
\left( \vec{S}_{1n} \cdot \vec{S}_{2n} \right) \left( \vec{S}_{1,n+1} \cdot \vec{S}_{2,n+1} \right) &\to - s(s+1) \left( \vec{l}(x) \right)^2 \\
\left( \vec{S}_{1n} \cdot \vec{S}_{1,n+1} \right) \left( \vec{S}_{2n} \cdot \vec{S}_{2,n+1} \right) &\to - s(s+1) \left( \vec{l}(x) \right)^2  - \frac{s^3 (s+1) \delta^2}{2} \left( \frac{\partial \vec{\varphi}}{\partial x} \right)^2 \\
\left( \vec{S}_{1n} \cdot \vec{S}_{2,n+1} \right) \left( \vec{S}_{2n} \cdot \vec{S}_{1,n+1} \right) &\to  - \frac{s^3 (s+1) \delta^2}{2} \left( \frac{\partial \vec{\varphi}}{\partial x} \right)^2
\end{split}
\ee
Finally, summing up all the terms from the Hamiltonian, 
\be
\begin{split}
&H=\frac{vg^2}{2}  \left( \vec{l}(x) \right)^2 + \frac{v}{2g^2} \left( \frac{\partial \vec{\varphi}}{\partial x} \right)^2, \\
\frac{vg^2}{2} = \beta_1 + \beta_2 - s & (s+1) (\beta_4 + \beta_5);~~~ \frac{v}{2g^2}= \frac{s^2 \delta^2}{2} \left( \beta_2 - \beta_3 - s(s+1) (\beta_5 + \beta_6) \right)
\end{split}
\ee

This effective Hamiltonian corresponds to the $O(3)$ non-linear $\sigma$-model with zero $\theta$-term; it describes at every site of the lattice a particle moving on a sphere $\left(\vec{ \varphi} (x) \right)^2 = 1$ with angular momentum $\vec{l}(x)$ \cite{Hamer:1979kx,Shankar:1990yq}. The angular momentum takes all possible integer values, $l(x)= 0,1, \cdots, \infty$. In the limit $g \gg 1$ the kinetic term $\frac{vg^2}{2}  \left( \vec{l}(x) \right)^2 $ dominates over the potential term $\frac{v}{2g^2} \left( \frac{\partial \vec{\varphi}}{\partial x} \right)^2$, and the ground state is obtained choosing the smallest possible value of $l(x)$ at every site. The first excited states are obtained by choosing the irreducible representation $l=1$ at one site and $l=0$ in the rest of the chain. Since this can be done at every site, there is a huge degeneracy, which is broken by the potential term. It delocalizes the $l=1$ excitations. The $3N$ degenerate first exited states become a band of $l=1$ magnons, separated from the ground state by a gap, which can be computed using perturbation theory. In \cite{Hamer:1979kx} this gap was computed up to 6th order in $\frac{1}{g^2}$. The first three terms read
\be
\Delta = vg \left( 1- \frac{2}{3 g^2} + \frac{0.074}{g^4} + O(\frac{1}{g^6}) \right)
\ee

In the weak coupling limit, a perturbative RG analysis \cite{Polyakov:1987ai} shows that the gap vanishes exponentially as 
\be
\Delta \simeq \frac{v}{g} e^{-2\pi/g}
\ee

\begin{figure}[!ht]
\begin{center}
\resizebox{!}{4.0cm}{\includegraphics{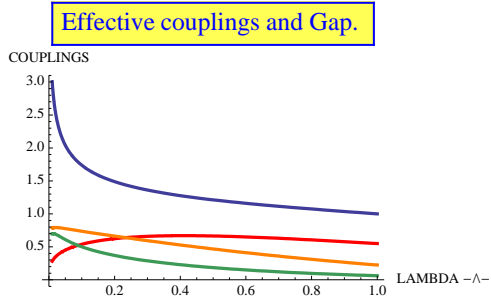}}
\caption{\label{effective} Values of effective coupling constant that define the $O(3)$ non-linear $\sigma$-model and minimum gap obtained from perturbation theory. Blue corresponds to the coupling constant $g$; Red corresponds to the effective velocity $v$; Orange corresponds to the minimum gap $\Delta$ in the strong coupling limit and green one to the gap in the weak coupling limit.}
\end{center}
\end{figure}


\section{Bosonization of a spin ladder}

In this section, we will apply another well-known tool from field theory, the bosonization of the spin degrees of freedom (see for instance \cite{Brezin:1989qv,Balents:1996la,Gogolin:1998qq,Delft:1998ek} and reference therein). This technique is complementary to the mapping to the sigma model, in the sense, that this method becomes exact when the coupling between spins in a rung of the ladder is weak while the one applied in the last section corresponds to the strong coupling limit. Nonetheless, we will see that the phase diagram covered by bosonization is in some sense richer than with the sigma model, giving the possibility to characterize different phases and to understand the transition between them. Also, it should be mentioned that bosonization techniques will allow us to have  a qualitative description of the behavior of the system but several non-universal, i.e. model dependent, quantities will appear in the calculations that should be matched with numerical techniques.

In the next lines, we will apply a method known as abelian bosonization keeping just the most relevant terms in the interactions. We will see that this procedure is good enough to give the correct physics that appears in spin ladders. Nonetheless, a more careful and detailed study of the ladder is probably needed. Several features have not been analyzed like the importance of marginal operators or the $SU(2)$ invariance is not explicitly kept in our derivation; for that propose, we suppose that more advance methods like nonabelian bosonization would be required (see for example \cite{Ludwig:1995yf}). 

In what follows, we would like to look at the physics of two leg Heisenberg spin-$1/2$ ladders. We will start studying the properties of two decoupled Heisenberg chains and then we will introduce a weak coupling between the chains along the rung and the plaquettes of the ladder. Hence, a one-dimensional Heisenberg spin chain is defined by the Hamiltonian,
\be
H=\sum_n \vec{S}_n \vec{S}_{n+1} = \sum_n  \left( S^x_n S^x_{n+1} +  S^y_n S^y_{n+1} \right) + \sum_n  S^z_n S^z_{n+1} = H_{xy} + H_{z}
\ee
with the usual spin operators, $[ S^{\alpha}_n, S^{\beta}_m]=i \delta_{n,m} \epsilon^{\alpha \beta \gamma} S^{\gamma}_n$, and $\{ \alpha, \beta, \gamma \} \in \{ x, y, z \}$. In the previous Hamiltonian, we have split the contribution that comes from the interaction in the $xy$-plane from the interaction in the $z$-spin direction, for a reason that will become immediately clear.

The first step to describe a spin chain in terms of bosons is to write the spin operators in terms of Jordan-Wigner fermions, i.e.
\be
\hat{a}_l = \left( \prod_{m<l} 2 S^z_m \right) \left( S^x_l -i S^y_l \right); ~~~~ \{\hat{a}_l, \hat{a}_m \}= \delta_{l,m} ~~ \{ \hat{a}_l, \hat{a}_m \}=0.
\ee
With this transformation the Hamiltonian is recast into
\be
H= \frac{1}{2} \sum_n \left( \hat{a}_n \hat{a}^{\dagger}_{n+1} + \hat{a}_{n+1} \hat{a}^{\dagger}_{n} \right) + \sum_n \left( \hat{a}^{\dagger}_n \hat{a}_n - \frac{1}{2} \right) \left( \hat{a}^{\dagger}_{n+1} \hat{a}_{n+1} - \frac{1}{2} \right)
\ee
From this equation we realize that the first part of the Hamiltonian, the one coming from the $xy$-interaction, is bilinear in fermion operators; hence, with a Fourier transformation, we can diagonalize $H_{xy}$. Moreover, taking the continuum limit, keeping the lattice spacing $\alpha$, and linearizing the Hamiltonian around the Fermi points ($k_F=\pm \frac{\pi}{2 \alpha}$), the first part of the Hamiltonian has two independent contributions due to left and right moving fermions,
\be
H_{xy}=i \alpha \int dx \left( \hat{a}^{\dagger}_L(x) \partial_x \hat{a}_L(x) - \hat{a}^{\dagger}_R(x) \partial_x \hat{a}_R(x) \right)
\ee
with
\be
\hat{a}(x)=\frac{\hat{a}_{n}}{\sqrt{\alpha}} = e^{ik_F x} \hat{a}_R(x) + e^{-ik_F x} \hat{a}_L(x)
\ee
Now, the crucial point made by bosonization is that the low energy description of this model is given by particle-hole excitations, i.e., it is characterized by excitations written in terms of bilinear fermionic operators, like the electron density, which fulfill bosonic commutation relations
\be
\begin{split}
\rho_R(x) = : \hat{a}^{\dagger}_R(x)  \hat{a}_R(x) :,& ~~~~ \rho_L(x) = : \hat{a}^{\dagger}_L(x)  \hat{a}_L(x) :, \\
[\rho_{\mu}(x), \rho_{\nu}(y)] =& \frac{i \delta_{\mu \nu}}{2 \pi} \partial_x \delta(x-y)
\end{split}
\ee
where $: :$ means normal ordering with respect to the ground state of the $H_{xy}$ Hamiltonian. With these definitions (and up to operators that modify the expectation values with logarithmic corrections \cite{Affleck:1998dv}), the Heisenberg Hamiltonian can be recast into,
\be
H = H_{xy} + H_{z} = \frac{\pi u}{2} \int dx \left( \frac{1}{g} \left( \rho_L(x) + \rho_R(x) \right)^2 + g \left( \rho_L(x) - \rho_R(x) \right)^2 \right)
\ee
with $u=\frac{\alpha}{\pi} \sqrt{(\pi -1)(\pi +3)}$, which corresponds to a renormalized Fermi velocity that always can be fixed to $u=1$ and $g=\sqrt{\frac{\pi -1}{\pi +3}}$; this value corresponds to a weak coupling calculation, comparing the field theory methods and Bethe ansatz results, it is known that in the (isotropic) Heisenberg model $g=\frac{1}{2}$.

With these transformations, the fermionic operators have a simple form in a bosonic language,

\be
\begin{split}
&\rho_R(x)  = \frac{1}{4 \pi}   \left(  \sqrt{g} -  \frac{1}{\sqrt{g}} \right) \partial_x \phi_L(x) + \frac{1}{4 \pi}  \left( \sqrt{g} +  \frac{1}{\sqrt{g}}  \right) \partial_x \phi_R(x)   \\
&\rho_L(x) =  \frac{1}{4 \pi}   \left(  \sqrt{g} + \frac{1}{\sqrt{g}} \right) \partial_x \phi_L(x) + \frac{1}{4 \pi}  \left(   \sqrt{g} - \frac{1}{\sqrt{g}}\right) \partial_x \phi_R(x) 
\end{split}
\ee
where the $\phi_{\mu}$  fields fulfill the following commutation relations, 
\be
[\phi_{\mu}(x), \partial_{y} \phi_{\nu}(y)]=2 \pi i \delta{(x-y)} \delta_{\mu, \nu}
\ee
and the Hamiltonian is recast into,
\be
H = H_{xy} + H_{z} = \frac{u}{8 \pi} \int dx \left(  \left( \partial_x \Phi(x) \right)^2 +  \left( \partial_x \Theta(x) \right)^2 \right)
\ee
with the bosonic field $\Phi(x) =  \left( \phi_L(x) + \phi_R(x) \right)$ and the dual one $\Theta(x)= \left( \phi_L(x) - \phi_R(x) \right)$.

Extending the relations between the bosonic and fermionic fields to the spin operators, the following equations can be found
\be
\begin{split}
S^z(x)&= \frac{\sqrt{g}}{2\pi} \partial_x \Phi(x) + \frac{(-1)^{x/\alpha}}{\pi \alpha} \sin{\left( \sqrt{g} \Phi(x) \right)} \\
S^+(x)&= \frac{(-1)^{x/\alpha}}{\sqrt{2 \pi \alpha}} \exp{\left( \frac{i \Theta(x)}{2 \sqrt{g}} \right)}
\end{split}
\ee

In the next lines, we will bosonize the two legs spin ladder, following the previous preliminary steps. Although, there is a huge literature on this subject (see for example \cite{Schulz:1986ph,Allen:1997gd,Strong:1994it} ), we will follow the methods used by Tsvelik and coauthors \cite{Tsvelik:1990kx,Shelton:1996la,Nersesyan:1997ek,Nersesyan:2003it} . In particular, we will see that perturbed field theory for the Heisenberg ladder has only one relevant perturbation which preserve the SU(2) symmetry and the leading SU(2)-invariant perturbation is of dimension 1.

For that propose, we introduce a new index $\sigma \in \{ 1, 2 \}$ for every spin operator which specifiy the chain of the ladder. The model, that we consider, is given by the Hamiltonian
\be
\begin{split}
H&= H_{\text{rung}}+H_{\text{leg}}+H_{\text{plaquette}} \\
 & =\beta_1 \sum_n \left( \vec{S}_{n,1}\cdot \vec{S}_{n,2} \right) + \beta_2 \sum_{n, \sigma} \left( \vec{S}_{n,\sigma}\cdot \vec{S}_{n+1,\sigma} \right) + \beta_4 \sum_n \left(\vec{S}_{n,1}\cdot \vec{S}_{n,2} \right) \left(\vec{S}_{n+1,1}\cdot \vec{S}_{n+1,2} \right) \\
& + \beta_5  \sum_n \left(\vec{S}_{n,1}\cdot \vec{S}_{n+1,1} \right) \left(\vec{S}_{n,2}\cdot \vec{S}_{n+1,2} \right) + \beta_6   \sum_n \left(\vec{S}_{n,1}\cdot \vec{S}_{n+1,2} \right) \left(\vec{S}_{n,2}\cdot \vec{S}_{n+1,1} \right)
\end{split}
\ee
The first term in the Hamiltonian is the Heisenberg interactions between the nearest neighbor spins in a rung of the ladder, the second term is the Heisenberg interaction between the nearest neighbor spins in a chain of the ladder; finally, the third, fourth and fifth interactions describe four-body spin interactions within in a plaquette in the ladder.

\begin{figure}[!ht]
\begin{center}
\resizebox{!}{3.0cm}{\includegraphics{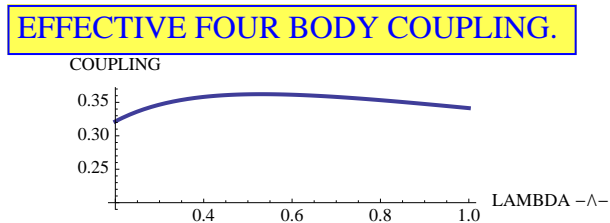}}
\caption{\label{plaquette} Estimation of the effective plaquette interaction in the ladder coming from the multipartite valence bond state. In the plot, it is presented the sum of the absolute values of four body interactions $\left( |\beta_4|+|\beta_5|+|\beta_6| \right)$ over the interaction along the rung $\beta_1$}
\end{center}
\end{figure}

The second term in the Hamiltonian is the one that we have just obtained and its bosonized expression reads
\be
H_{\text{leg}}=\frac{u\beta_2 }{8 \pi} \sum_{\sigma} \int dx \left(  \left( \partial_x \Phi_{\sigma}(x) \right)^2 +  \left( \partial_x \Theta_{\sigma}(x) \right)^2 \right)
\ee
The first term in the ladder Hamiltonian, using the definitions of the spin operators in terms of bosonic field and keeping the most relevant terms, can be recast into,
\be
H_{\text{rung}}= \frac{\beta_1}{2\pi \alpha} \int dx \Big[ \cos{\left( \sqrt{g} \left( \Phi_1(x) - \Phi_2(x) \right) \right) }- \cos{ \left( \sqrt{g} \left(\Phi_1(x) + \Phi_2(x) \right)  \right)}+ 2 \cos{\left(\frac{\Theta_1(x) - \Theta_2(x)}{2\sqrt{g}} \right) } \Big]
\ee 
For the plaquette interactions, it can be seen that can be mapped to a similar bosonic interaction with an effective four body $\beta_{\text{eff}}$, that depends on the values of $\beta_4$, $\beta_5$ and $\beta_6$ and non universal coefficients like the lattice spacing $\alpha$, hence, it is actual value should be fixed comparing the numerical results and field theory methods \cite{Mueller:2002zr,Lauchli:2003qq},
\be
H_{\text{plaquette}}=\frac{\beta_{\text{eff}}}{2 \pi \alpha} \int dx  \Big[ \cos{\left( \sqrt{g} \left( \Phi_1(x) - \Phi_2(x) \right) \right) } + \cos{ \left( \sqrt{g} \left(\Phi_1(x) + \Phi_2(x) \right)  \right)} \Big]
\ee
Introducing the linear combinations of the fields $\Phi_1$ and $\Phi_2$:
\be
\Phi_{\pm} = \frac{\Phi_1 \pm \Phi_2}{\sqrt{2}}
\ee
the $\Phi_+$ decouples from the $\Phi_-$ contributions and the total Hamiltonian can be seen as the sum of two independent Heisenberg chains,
\be
\begin{split}
H_+ &=\frac{u\beta_2 }{8 \pi}  \int dx \left(  \left( \partial_x \Phi_{+}(x) \right)^2 +  \left( \partial_x \Theta_{+}(x) \right)^2 \right) + \frac{\beta_{\text{eff}}-\beta_1}{2 \pi \alpha} \cos{\left( \Phi_+\right)} \\
H_- &=\frac{u\beta_2 }{8 \pi}  \int dx \left(  \left( \partial_x \Phi_{-}(x) \right)^2 +  \left( \partial_x \Theta_{-}(x) \right)^2 \right) + \frac{\beta_{\text{eff}}+\beta_1}{2 \pi \alpha} \cos{\left( \Phi_-\right)} + \frac{\beta_1}{\pi \alpha} \cos{\left( \Theta_-\right)}
\end{split}
\ee
At this point, we can refermionize these models introducing a spinless Dirac fermion
\be
\tilde{a}_{R,L}(x) = \frac{1}{\sqrt{2\pi \alpha}} \exp{\left( \pm i \phi_{+; R,L}(x)\right)}
\ee
then, the Hamiltonian $H_+$ recast into a free massive fermion theory,
\be
H_+=i v_{\text{eff}} \int dx \left( \tilde{a}^{\dagger}_L \partial_x \tilde{a}_L - \tilde{a}^{\dagger}_R \partial_x \tilde{a}_R \right) + i m_t \int dx  \left( \tilde{a}^{\dagger}_L \tilde{a}_R - \tilde{a}^{\dagger}_R \tilde{a}_L \right) 
\ee
with $m_t=\frac{\beta_1 -\beta_{\text{eff}}}{2 \pi}$
The Hamiltonian $H_-$ also appears as a free massive fermion theory but including a pairing interaction,
\be
\begin{split}
&\tilde{c}_{R,L}(x) = \frac{1}{\sqrt{2\pi \alpha}} \exp{\left( \pm i \phi_{+; R,L}(x)\right)} \\
&H_-=i v_{\text{eff}} \int dx \left( \tilde{c}^{\dagger}_L \partial_x \tilde{c}_L - \tilde{c}^{\dagger}_R \partial_x \tilde{c}_R \right) + i \frac{\beta_1 +\beta_{\text{eff}}}{2 \pi \alpha}\int dx  \left( \tilde{c}^{\dagger}_R \tilde{c}_L - \tilde{c}^{\dagger}_L \tilde{c}_R \right) +\frac{i\beta_1}{\pi \alpha} \left( \tilde{c}^{\dagger}_R \tilde{c}^{\dagger}_L - \tilde{c}_L \tilde{c}_R \right)
\end{split}
\ee
Finally, defining the (real) Majorana fermions,
\be
\check{a}^x_{\mu}=\frac{\tilde{a}_{\mu}+\tilde{a}^{\dagger}_{\mu}}{\sqrt{2}},~~~ \check{a}^y_{\mu}=\frac{\tilde{a}_{\mu}-\tilde{a}^{\dagger}_{\mu}}{i\sqrt{2}},~~~\check{a}^z_{\mu}=\frac{\tilde{c}_{\mu}+\tilde{c}^{\dagger}_{\mu}}{\sqrt{2}},~~~ \check{a}^0_{\mu}=\frac{\tilde{c}_{\mu}-\tilde{c}^{\dagger}_{\mu}}{i\sqrt{2}}, ~~~(\mu \in{L,R}) \\
\ee
the total Hamiltonian is recast into the sum of four independent free massive majorana fermions,
\be
H= H_+  + H_- = \sum_{\mu=\{ x,y,z \}}H_{m_t}[\check{a}^{\mu}] +H_{m_s}[\check{a}^{0}] 
\ee
with $m_s=-\frac{3\beta_1 +\beta_{\text{eff}}}{2 \pi}$ and
\be
H_m[\check{a}]= \frac{i v_{\text{eff}}}{2} \int dx \left( \check{a}_L \partial_x \check{a}_L - \check{a}_R \partial_x \check{a}_R \right) + i m \int dx  \left( \check{a}_L \check{a}_R \right) 
\ee
From the above formulas, one readily obtains the line where the triplet mass vanishes, i.e. $m_t  \to 0$, when $\frac{\beta_{\text{eff}}}{ \beta_1} \sim O(1)$ (see Fig.\ref{plaquette}). This transition belongs to the universality class of the critical, exactly integrable, $S=1$ spin chain (Takhtajan-Babujian point \cite{Takhtajan:1982fk,Babujian:1983uq}) with the cental charge $c=3/2$. Based on the fact that the critical line is specified in terms of three massless majoranas, the $SU(2)_2$ WZNW model is the universality class describing this line. The gap, which is the mass of the majorana triplets, opens linearly as one deviates form criticality. Owing to the $SU(2)$ symmetry of the model no other perturbations than mass terms of majoranas are allowed.

\section{Numerical results}

In this section, we study the ladder system resulting from the mapping defined by Eq.\ (\ref{hamlad}) by numerical means. The Hamiltonian of this system depends on two parameters $\phi \in [\pi/6, 2 \pi/3 ] $ and $\Lambda \in [0,1]$ which fix the six parameters of Eqs.\ (\ref{ladpar}). 

In the following, we present results concerning the exact diagonalization of ladders up to 28 spins or length $L = 14$ for different values of $\Lambda$ with open boundary conditions and using DMRG calculations. Several points seem to be remarkable. First, there is a region of criticality around the point $\Lambda \to 1$; as the system approaches this region, it can be described by two decoupled Heisenberg chains, a critical system. Second, the first excited state is three-fold degenerate. Third, in Fig. \ref{dmrg}, we plot the minimum gap as a function of the parameters $(\phi, \Lambda)$. From the plot, it seems that the angle $\phi$ is an irrelevant parameter at least for a substantial part of the phase diagram $(\phi, \Lambda)$. Nevertheless, there are special regions that require a careful analysis. 

\begin{figure}[!ht]
\begin{center}
\resizebox{!}{4.5cm}{\includegraphics{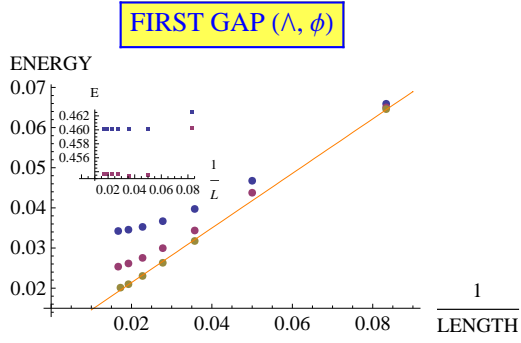}}
\caption{\label{gapdmrg}Results of DMRG calculations of first energy gap as a function of the length and the number of states kept in the simulations, with $(\Lambda, \phi)=(1, 1.15)$ in the main plot and $(\frac{2}{3}, \frac{\pi}{2})$ in the inset. Blue points correspond to a number of states $m=40$, purple points to $m=60$ and golden points to $m=100$. In the main plot the golden points follow a linear behaviour expected in a critical theory.}
\end{center}
\end{figure}

To achieve some knowledge about the behavior of the system in the thermodynamical limit, we used the DMRG method \cite{White:1992fs,Schollwoeck:2004pz} to find the system ground state and the first excited state energy. We sample the two-dimensional parameter space with one hundred points, and for ladders of up to $L=60$ rungs, retaining $m=40$ states to compute the gap of the system (we checked convergence of the results till $m=100$ for some typical points). Typical results are shown in Fig. \ref{gapdmrg}. These results have been compared with exact diagonalization calculations for small instances of the problem, e.g.\ $L=14$. The complete phase diagram for $L=60, m=40$ is displayed in Fig. \ref{dmrg}.

From the phase diagram it is clear that the gap is closing for $\Lambda \to 1$. Due to the closing gap, the correlations in the ground state become stronger and thus the numerical simulations require more and more resources. We moved to a ladder with $L=100$, retaining $m=60$. In Fig. \ref{gapdmrg} the gap values for $\Lambda=1$ and $\phi =1.15$ are shown.  In the inset, it can be also seen the results in the regime $\Lambda= \frac{2}{3}$ and $\phi =\frac{\pi}{2}$ which signal a different behavior.

Exact diagonalisation is used here to provide exact information about the low energy spectrum of the system with the size up to N=28 spins and to benchmark the DMRG results. Exact diagonalization by construction is an unbiased and reliable method but its main drawback lies on the limitation of the systems that can be analyzed due to the exponential growth of the Hilbert space with the system size. The developed code utilizes the PETSc \cite{:yq} and SLEPc \cite{:rt}  libraries and is run on a Blue Gene/P supercomputer utilising up to 2048 processing cores.

\begin{figure}[!ht]
\begin{center}
\resizebox{!}{6.0cm}{\includegraphics{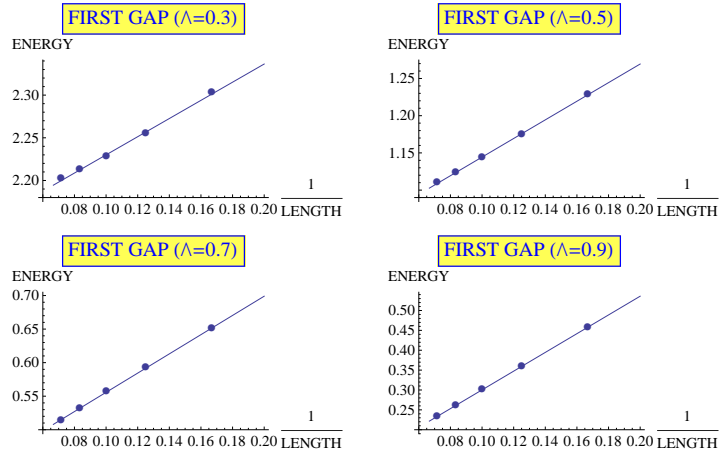}}
\caption{\label{firstgap}Results of exact diagonalization of first energy gap as a function of the length and $\Lambda$ for length up to $L=14$. All the plots show a clear linear dependence of the gap with the inverse of the length of the ladder. The lines correpond to the best linear fits and are just to guide the eye}
\end{center}
\end{figure}

The question that we try to solve with exact diagonalization is the behavior of the minimum gap of the system with the length and the perturbation $\Lambda$. We already got some a priori knowledge about the possible scenarios that can appear due to the field theoretical analysis. In the case that the system is gapless, in the thermodynamical limit, the minimum gap should present a linear dependence with the system size as it is shown in the Fig. \ref{firstgap}. 

\begin{figure}[!ht]
\begin{center}
\resizebox{!}{4.0cm}{\includegraphics{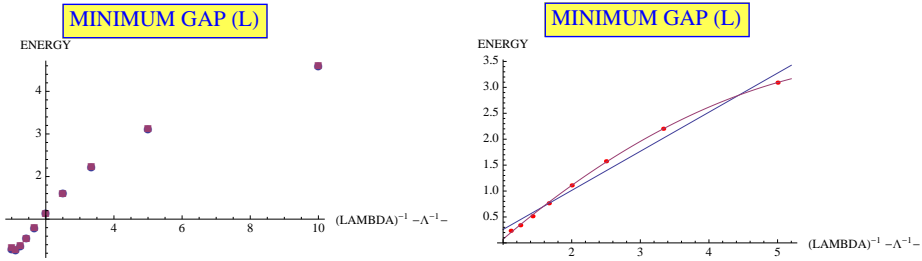}}
\caption{\label{gaplam} Results of exact diagonalization of first energy gap as a function of the length and $\Lambda$. On the left side, it is shown the plots for system sizes $L=\{12, 14\}$. One the right side, it is plotted a closer look of the results for $L=14$. The blue line correponds to the best linear fit, the red line corresponds to the best quadratic fit; they are just to guide the eye but it seems that the plots does not show a linear dependence of the gap with the perturbation $\Lambda$.}
\end{center}
\end{figure}

Another convincing argument about the gapless system is to show how the gap scales with the perturbation $\Lambda$. Now, we know that perturbed conformal fields theory shows that there is only one relevant perturbation which preserves the $SU(2)$ symmetry of the system with scaling dimension one, which means that close to the $\Lambda =1$, where we have two decoupled Heisenberg chains, the gap of the system should open linearly if the system remains gapped in the thermodynamical limit. In Fig. \ref{gaplam}, we have plotted the first energy gap as a function of $\Lambda^{-1}$, there, we can see that the plots does not show a linear dependence of the gap with the perturbation $\Lambda$.

\section{\label{sec:Neutron-scattering-background}Neutron scattering using the 2D multipartite valence bond state}

This section is dedicated to the study of the real-world relevance of the theoretical model. To this aim, we compare the two-point equal-time correlation function obtained from the theoretical predictions of our anti-ferromagnetic model and the linear spin wave theory with data from neutron scattering experiments of cuprates in the anti-ferromagnet insulator phase.

In the context of high-$T_c$ superconductivity, the investigation of the different competing phases of a potentially superconducting material and the transition between them is of strong interest. In the limit of low temperature and no doping, a very simple structure is dominant: the N\'eel order, the signature of an anti-ferromagnetic long-range ordered state. However, this phase is not the interesting one. Changing the environmental conditions or doping might cause a phase change, in some cases leading to superconductivity. These \emph{normal phases} do not show long range order and have also other interesting structural properties. Among these is the formation of \emph{stripes}, whose occurrence is conjectured by experimentalists to be linked to high-$T_c$ superconductivity \cite{Tranquada:1995qd,Tranquada:1997hl,Kivelson:2003qf,Buschow:2003qr,Bennemann:2008rc}. In this section we want to investigate the multipartite valence bond state from this point of view. Since it is lacking long range order, it might be a candidate for the description of normal states of a wide range of materials. To give an example, we might be able to model cuprates, which are assembled from spin-$1/2$ sites, but might exhibit -- in a low-energy sector of the normal phase -- the behaviour of a substrate made of spin-$1$ sites. Another example is the nickelates which are actually assembled directly from spin-$1$ sites. In this case, we might be able to grasp even the full theory, not only an effective one.

The investigation carried out in this section is the numerical analysis of a hypothetical neutron scattering experiment, performed on the MPVBS. It is then compared to experimental data as well as predictions of other theoretical models. Our focus will be on the short-range entanglement features exhibited by the experiment (and the predictions), because these features enter the construction of the MPVBS on a very fundamental level.

The aim of and the essential idea behind a scattering experiment is to find out which degrees of freedom (to absorb energy and momentum, etc., from the incident scattered beam) a target system has, and from this to draw conclusions about the internal structure of the target. In principle, a theoretical model derived from first principles should provide us with these degrees of freedom and therefore with the possible outcomes of the interaction of beam and target. 

The central quantity for our following analysis is the experimentally and theoretically accessible \emph{differential cross section}
\[
\left(\frac{d\sigma}{d\Omega}\right)_{\mbox{ground state}\rightarrow\mbox{any}}.
\]
We assume a setting where the incoming and outgoing beams are waves of definite momentum.  We are not interested in the final state of the target system and choose the ground state of the anti-ferromagnet as our target. The fact that we are not interested in the final state liberates us from the task to determine the exact Hamiltonian; when traced out, it is not important. In rotationally invariant states we obtain the formula
\[
\left(\frac{d\sigma}{d\Omega}\right)_{\mbox{ground state}\rightarrow\mbox{any}}
=\sum_{\mathbf{r}_{i},\mathbf{r}_{j}}\exp\left[-i\boldsymbol{\mathbf{\kappa}}\left(\mathbf{r}_{i}-\mathbf{r}_{j}\right)\right]\left\langle \lambda|S_{z,i}S_{z,j}|\lambda\right\rangle,
\]
which implies a relation between the two-point correlation function and the differential cross section. For further references and details, we refer to e.g.\ \cite{Lovesey:1986yq}. The correlation functions are accessible in calculations involving the MPVBS. In this section and using this relation, we will now make further use of these (numerically accessible) quantities and hence be able to compare them to cross sections of other models, techniques and to experimental data. One of these alternatives techniques is the linear spin wave theory.

\subsection{The linear spin wave approach in an anti-ferromagnet\label{sub:Spin-waves} and alternative theories}

Assuming that we have already found the right model for the description of a solid state system, a big challenge remains: the problem of performing the quantum mechanical calculation in the thermodynamic limit, i.e., for objects of macroscopic size that are observed in the laboratory and usually consist of $\sim 10^{23}$ particles. In the context of scattering experiments, but not only there, a good way to reduce the complexity of the macroscopic system is to make certain a priori assumptions about its degrees of freedom. One of the conventional assumptions that can be made and that will be used in this paper as a reference is the idea that the excitations of \emph{spin waves} are the essential events that take energy and momentum from the incident beam of scattered particles.

If we assume that the system is properly described by the Heisenberg Hamiltonian on a square lattice 
\[
H=J\sum_{\left\langle \mathbf{j},\mathbf{k}\right\rangle }\mathbf{S}_{\mathbf{j}} \cdot \mathbf{S}_{\mathbf{k}},\quad J>0,
\]
where $\left\langle \mathbf{j},\mathbf{k}\right\rangle $ is a sum over nearest neighbors, then in the corresponding \emph{classical} ground state every two neighbors would be aligned in parallel, but with opposite orientation. This leaves us with two sub-lattices (I and II) with equally oriented spins, one sub-lattice pointing up and one down. This ordering is called the N\'eel ordering and the corresponding state with this ordering is called the N\'eel state. The deviations from this order can be described in terms of modes with definite wavelength. The quantization of these modes leads to another quantum theory of the state. The linearized version of this theory is the traditional description of the quantum mechanical anti-ferromagnet: linear spin wave theory.

Hulthen \cite{Hulthen:1936kx} carried through the quantization of these spin waves, which subsequently were used by Anderson \cite{Anderson:1952th} in his approximate theory of the quantum mechanical anti-ferromagnetic ground state. They finally obtain the proportionality
\[
\sum_{\mathbf{j},\mathbf{l},\tiny{\mbox{ both from }}I,II}\exp\left[-i\mathbf{k}\left(\mathbf{j}-\mathbf{l}\right)\right]\left\langle S_{\mathbf{j},x}S_{\mathbf{l},x}\right\rangle
\propto\sqrt{\left(1-\gamma_{\mathbf{k}}\right)/\left(1+\gamma_{\mathbf{k}}\right)},
\]
where $\gamma_{\mathbf{k}}=\sum_{i}\cos k_{i}$, i.e., $\sqrt{\left(1-\gamma_{\mathbf{k}}\right)/\left(1+\gamma_{\mathbf{k}}\right)}$ is proportional to the differential  scattering cross section into the direction of $\mathbf{k}$.

Another approach worth mentioning is the superposition of the classical ground state of the anti-ferromagnet -- the N\'eel state -- with fluctuating valence bonds. This approach is a construction of (an approximation of) a ground state that offers a qualitative understanding but fails to make the right predictions in certain regions of $\mathbf{Q}$-space. We will not elaborate on this approach here, but account for its predictions later, if appropriate.

As it has already been pointed out in \cite{Christensen:2007it}, there are many more models and techniques to make predictions beyond linear spin wave theory and the multipartite valence bond state. A very important technique is, of course, quantum Monte-Carlo, which gives the right answers (or at least being very close), as indicated by simulations at certain points of interest in $\mathbf{Q}$-space which were performed and discussed in \cite{Christensen:2007it}. The approximate correctness of quantum Monte-Carlo simulations for this model gives confidence that the basic assumptions about the model are justified and we can concentrate on finding its solutions. However, the quantum Monte-Carlo simulations are not suited to understand the features of the solution \emph{qualitatively} and are hence unsatisfactory. We also stress again that the Heisenberg Hamiltonian is not the \emph{optimal} choice either, since it is known that plaquette interactions take place in the real-world anti-ferromagnetic substrates \cite{Coldea:2001hl}. However, the family of Heisenberg-like Hamiltonians is large and includes alternative anti-ferromagnetic Hamiltonians; also the parent Hamiltonian of the multipartite valence bond state belongs to this class. 

\subsection{The experiment}

Recently, the various scattering cross sections of a two-dimensional square-lattice anti-ferromagnet have been measured experimentally in great detail \cite{Christensen:2007it}, offering a great opportunity to test a wide range of theoretical models.

The substance to be measured was a single crystal of CFTD {[}Cu(DCOO)$_{2}\cdot$4D$_{2}$O], with lattice constants $a=8.113\AA$, $b=8.119\AA$ and $c=12.45\AA$. In an effective model of CFTD, the essential sites (i.e., those that are responsible for the properties of the substance) are given by the electrons in the $d$-shells of the copper ions, naturally being spin-1/2 particles. In the ionic picture, the electrons are strongly localized and there is almost no coupling between two layers in the $c$-direction. Moreover, the effective interactions between these sites cause the crystal to be anti-ferromagnetic. Hence CFTD is an almost perfect realization of a two-dimensional square lattice Heisenberg anti-ferromagnet.

In the experiment, several assumptions were tested. (i) One assumption is the idea that the quantum anti-ferromagnetic ground state is essentially the classical counterpart with minor quantum corrections, which -- once found -- are moreover well understandable conceptionally. This assumption expresses the hope that the quantum anti-ferromagnetic ground state can be understood qualitatively. (ii) The second assumption that was tested is the idea that spin waves are the essential degrees of freedom of excitation in the anti-ferromagnetic ground state.

We furthermore notice that both linear spin wave theory and the experiment assume the interaction to be described by the Heisenberg Hamiltonian. This assumption makes these models different from the MPVBS, because its parent Hamiltonian contains plaquette terms.

\subsection{Comparison of theoretical predictions with experimental data}

We compare the experimental data with the predictions of the linear spin wave theory and the two-dimensional multipartite valence bond state. In both cases, the differential cross section is proportional to the Fourier transform of the two point function. The calculation of this two point function is straightforward but -- without further simplifications -- naturally becomes more and more tedious as the system size grows. In the case of the multipartite valence bond state, we did not use any further assumptions about the degrees of freedom but calculated the two-point function directly. In fact, we fix the parameters $(\Lambda, \phi)$ that defined the MPBVS to the values $(\frac{1}{3},\frac{\pi}{2})$. We have performed the calculations up to a grid size of $8\times8$, and we use the results of this lattice for the analysis. Finite size effects are an issue in such a small lattice and will be most notable for wave vectors of large magnitude, i.e., in the center of the $\mathbf{Q}$-space, which is given by the interval $\left(0,2\pi\right)\times\left(0,2\pi\right)$.

\begin{figure}[ht]
\includegraphics[width=5cm]{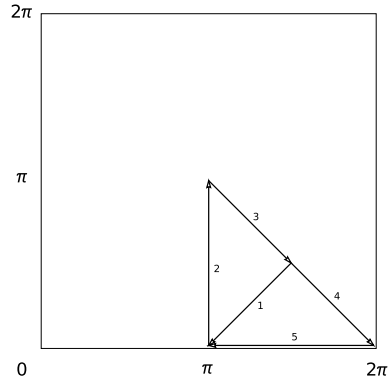}
\caption{\label{fig:path}The path to follow.}
\end{figure}

We will now discuss the cross sections in greater detail. Initially, the cross section is given as a function
\[
\left(0,2\pi\right)\times\left(0,2\pi\right)\ni\mathbf{Q}
\mapsto\left(\frac{d\sigma}{d\Omega}\right)_{\mbox{GS}\rightarrow\mbox{any}}
\left(\mathbf{Q}\right).
\]
For convenience, we will follow this function on a path -- being depicted in Fig.\ \ref{fig:path} -- that crosses several interesting regions of $\mathbf{\mathbf{Q}}$-space; the reasons for these regions being interesting will be explained later. This way we arrive at a two-dimensional plot of scattering cross section versus length on the path, which is more accessible than a three-dimensional version. The shape of the path is motivated by experimental work \cite{Christensen:2007it} and the data provided therein.

\begin{figure}[t]
\includegraphics[width=0.9\columnwidth]{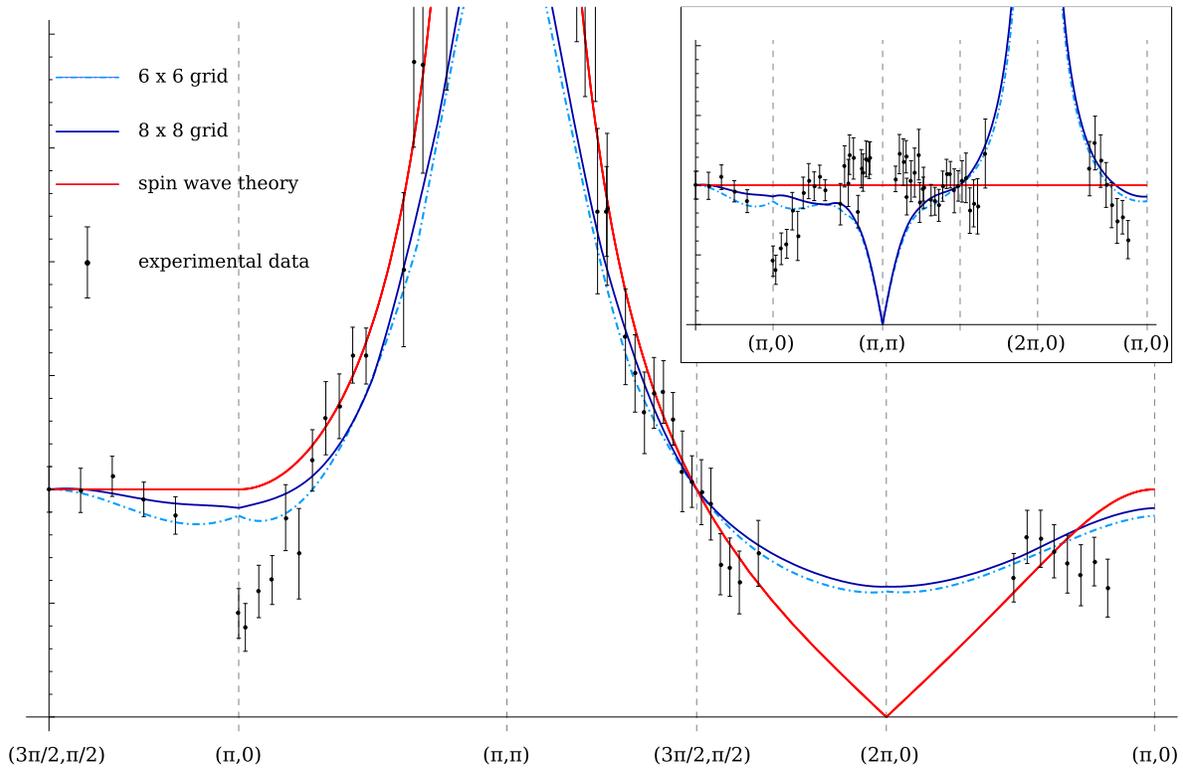}
\caption{\label{fig:CrossSecOne} Cross sections along the path.}
Features of the predictions of linear spin wave model, the multipartite valence bond state and the experimental data. The inset shows the same data, normalized such that the value of the LSWT equals one everywhere. G. Aeppli's group provided the experimental data published in \cite{Christensen:2007it}.
\end{figure}

Our discussion starts with the raw qualitative features of the predicted and experimental cross sections using Fig. \ref{fig:CrossSecOne}. In this plot the cross sections of the linear spin wave model, the multipartite valence bond state and the experiment along the path are given. The units are arbitrary and all curves and data points are normalized to give the best possible match with the experiment overall.

We see that LSWT predicts the experimental data with excellent quality on a crude scale. The singularity at $\left(\pi,\pi\right)$ is predicted properly. Moreover, there are no \emph{huge} qualitative differences to be seen, although the cross section is false in some regions, showing a slightly wrong tendency there. Most interestingly, in the first part of patch $2$ of the path the cross section is generally too high, this feature being reflected again in the last part of patch $5$, which covers a similar region. A comparison of LSWT with the experimental data in this region shows that close to the point $\left(2\pi,0\right)$ LSWT predicts a too small scattering cross section; and close to $\left(\pi,0\right)$ the prediction is too large. The error of the prediction shows an almost linear relation along patch $5$. This feature is closely linked to the prediction of linear spin wave theory that at the point $\left(2\pi,0\right)$ the cross section should be zero. Considering the data, this is most likely a qualitatively wrong statement about the cross section, based on wrong assumptions about the degrees of freedom available and hence about the electron structure at these length scales. Generally, we observe that the cross section in the central region of the wave vectors, i.e., close to the point $\left(\pi,\pi\right)$ are more precise in their predictions than those at the zone boundary, the latter belonging to wave vectors resolving features of the electron structure of small scale; a typical length being around the size of a small multiple of the distance Cu-Cu in the lattice.

Like the LSWT, the MPVBS follows the experiment very well, considering the finite size effects of the simulation. Please note that there are two regions of the plot that we should consider separately in the context of our simulation of the MPVBS. Finite size effects divide the zone into two regions of different precision: like in the case of LSWT we have to distinguish between wave vectors in central and peripheral regions, the latter being more precise than the former numerically, since finite size effects have a great impact in the central region.

The first region is represented by the patches $1,2,3$ of the path. The singularity at $\left(\pi,\pi\right)$ is not predicted. We believe that the deviations in this region are mostly due to the finite size of the lattice that was used to perform our calculations, and not due to the qualitative features of the MPVBS. However, we were unable to prove this, since direct calculations with larger lattices have so far not been feasible, and in principle one should go to the thermodynamic limit. The patches $4$ and $5$ of the path belong to wave vectors at or close to the boundary of the zone and hence are not strongly affected by the small lattice size. Please note that this is the region of the most interesting deviations of LSWT from the experiment. Here, one qualitative difference between LSWT and the MPVBS is apparent; at the point $\left(2\pi,0\right)$ the LSWT scattering cross section vanishes completely and becomes comparatively small already in a comparatively large region near this point. The MPVBS shows a completely different behavior; the scattering cross section stays finite. Unfortunately the experiment does not provide data in this region.

As has been pointed out before, the patches $4$ and $5$ of the path are most expressive, but also the region where patch $1$ and $2$ meet, the finite lattice size does not have a strong impact. In the latter region, we see that LSWT predicts the experimental data with an error of $100\%$. The MPVBS, however, shows the right tendency. The predicted value is less than that of LSWT. The comparison of the qualitative behavior of the LSWT, the MPVBS and the experimental data is possible in an optimal way in a plot that shows the data from Fig.\ \ref{fig:CrossSecOne} normalized to the predictions of LSWT. Here the tendency in the patches $4$ and $5$ is most apparent. Please notice again that the patches $1,2$ and $3$ are severely affected by the finite size of the lattice used for computations.

What is the significance of patch $5$ and the meeting points of patches $1$ and $2$? The cross section in this region of the zone reflects the electron structure at short distances of the lattice. As has been conjectured already in \cite{Christensen:2007it}, the entanglement features on this length scale in the Heisenberg anti-ferromagnet and hence also in CFTD might be completely different from those of the N\'eel state and the spin wave excitations. On the one hand, the N\'eel state is a classical state with no entanglement and also the excitable degrees of freedom, the magnons, show the entanglement structure of collective modes of excitations of harmonic oscillators.

On the other hand, the MPVBS is constructed to have the entanglement structure of (two-electron) singlets, established between nearest neighbors. This assumption about the electron structure obviously allows for better predictions at the small length scale. An assumption similar to the construction of the MPVBS is made when we start with the N\'eel state as the ground state but allow for a superposition with resonating valence bonds, in the ground state as well in its modes of excitation. As shown in \cite{Christensen:2007it}, this construction gives a very good match at point $\left(\pi,0\right)$, but results at the same time in a prediction that is far too low -- though qualitatively right -- when approaching the point $\left(2\pi,0\right)$. This wave vector pointing in the diagonal direction of the lattice, where no valence bonds are assumed in the construction. Obviously some degrees of freedom are missing in this description, which again are included in the MPVBS. Please note in this context that the effective interactions in CFTD arise under mediation by formate groups in between the copper ions, a situation which is considered and reflected in the construction of the multipartite valence bond state.

\section{Conclusions}

In this paper we carried out several approaches for the analysis of the multi partite valence bond state described in Ref.~~\cite{Rico:2008rm}. This state is based on a local tensor description that enforces important physical symmetries. It can hence be interpreted as being a two-dimensional generalization of one of the most frequently studied states with such a local tensor structure, the ground state of the Affleck-Kennedy-Lieb-Tasaki (\emph{AKLT}) model \cite{Affleck:1987cm,Affleck:1987cy}, a spin-1 chain with a Heisenberg-like Hamiltonian. Moreover, it can be seen as a complementary generalization to their extension to higher dimensions.

Our main goal in this paper was to analyze the behavior of the two point correlation functions of the multipartite valence bond state model and to compare it with the classical linear spin wave theory (LSW) and experimental data.

To achieve this goal, we proceeded in two directions: first, we applied a mapping to a quantum ladder problem, in order to discern between an algebraic and an exponential decay of the correlations. In this step, we use analytical tools (mapping to the non-linear sigma model and bosonization techniques) and numerical tools (exact diagonalization and variational methods). Another strategy that should be investigated in the future, and we leave open, it is the study of the equivalent two dimensional classical vertex model, its integrability properties and the relations with the Yang-Baxter equations.

Second, we calculated the equal time two point functions and compared it with both the results of LSW and data from a neutron scattering experiment on a Mott-Hubbard anti-ferromagnet insulator material. 

In the context of the description of two-dimensional solid states in the anti-ferromagnet insulator phase, our model can be seen as an alternative to the linear spin wave approach. In fact, the scattering cross sections of both models differ qualitatively in important regions. These regions correspond to structures that the LSW approximation fails to describe properly and that are conjectured to correspond to important qualitative features of the electron structure in the actual solid. More precisely, these features are believed to be related to short-range entanglement structures of the electrons in the solid and hence reflect a design goal of the generalized AKLT model that we give.

Moreover, the multi-partite valence bond state offers a new and unusual constructive approach for toy models in solid state physics, providing unusual points of view on the structure of high-$T_c$ superconductivity.

\section{ACKNOWLEDGEMENTS}

This work was supported by networks DARPA, EMALI, EUROSQIP, FWF, MURI, NAMEQUAM, OLAQUI and SCALA and developed using the DMRG code released within the PwP project (http://www.dmrg.it). J.V. and N.M. acknowledge the support of SFI through the PIYRA and the computing resources of ICHEC. E.R. would like to thank The Hebrew University of Jerusalem for their hospitality during the winter school 2008 on Condensed Matter and Quantum Information. N.M and E.R. also would like to thank the Ecole de Physique Les Houches for their hospitality during the summer school 2008 on "Exact methods in low-dimensional Statistical Physics and Quantum Computing", where part of this work was done. The authors thank I. Affleck, M.A. Martin-Delgado, G. Sierra and F. Verstraete for useful and constructive comments and G. Aeppli's group for providing their experimental data on neutron scattering.

\bibliographystyle{unsrt}

\end{document}